\documentclass{ifacconf}

\usepackage{graphicx}      
\usepackage{natbib}        
\usepackage{graphicx}
\usepackage{graphics}
\usepackage{amsmath}
\usepackage{amssymb}
\usepackage{color}
\usepackage[noend]{algpseudocode}
\usepackage{algorithm}
\usepackage{enumitem}
\usepackage{epstopdf}
\usepackage{algorithmicx}

\begin{document}
\begin{frontmatter}

\title{Encrypted Price-based Market Mechanism for Optimal Load Frequency Control\thanksref{footnoteinfo}} 

\thanks[footnoteinfo]{This work is supported by NSF Award 1944318.}

\author[First]{Jihoon Suh} 
\author[First]{Takashi Tanaka} 

\address[First]{Department of Aerospace Engineering and Engineering Mechanics, The University of Texas at Austin \\ (e-mail: jihoonsuh@utexas.edu, ttanaka@utexas.edu)}

\begin{abstract}                
The global trend of energy deregulation has led to the market mechanism replacing some functionality of load frequency control (LFC). Accordingly, information exchange among participating generators and the market operator plays a crucial role in optimizing social utility. However, privacy has been an equally pressing concern in such settings. This conflict between individuals' privacy and social utility has been a long-standing challenge in market mechanism literature as well as in Cyber-Physical Systems (CPSs). In this paper, we propose a novel encrypted market architecture that leverages a hybrid encryption method and two-party computation protocols, enabling the secure synthesis and implementation of an optimal price-based market mechanism. This work spotlights the importance of secure and efficient outsourcing of controller synthesis, which is a critical element within the proposed framework. A two-area LFC model is used to conduct a case study.
\end{abstract}



\end{frontmatter}

\section{Introduction}\label{intro}
\vspace{-2ex}
Load frequency control (LFC), also known as automatic generation control (AGC), maintains the frequency of a synchronous power grid near the nominal operating point (typically $50$Hz or $60$Hz) by matching the generation level with consumption in a real-time manner.
In deregulated power systems in which generators are operated by a large number of private profit-seeking firms, LFC often needs to be performed through a market mechanism (\cite{berger1989real}). 
Roughly speaking, the power system operator (or the Independent System Operator, ISO) raises the real-time electricity price when consumption exceeds generation to incentivize the generators to increase the output, and does the opposite when generation exceeds consumption until the frequency deviation is eliminated.\footnote{In reality, multiple markets (such as day-ahead and real-time markets) coexist to operate a single grid. In this paper, we only consider a simplified market model as described in Section \ref{prelim}.} 
\vspace{-3ex}

From the ISO's perspective, this is a feedback control problem in which a price mechanism (control law) must be designed to achieve an effective frequency regulation. However, this is often a challenging problem as the ISO typically has very limited information about the market participant's \emph{type}.\footnote{In market design theory, \emph{type} refers to a collection of market participant's private information describing its preferences and constraints.} For example, generators' price-response behaviors are hard to predict as they depend on the physical properties of the generation facility (e.g., hydro turbines can react faster than combustion turbines) as well as their generation cost functions, and such information is not known to the ISO \emph{a priori}. In principle, such a limitation can be removed if all the market participants (generators) reveal their private information to the ISO prior to the market operation, and the ISO designs an effective price mechanism based on the full knowledge of the market. Unfortunately, this is not a realistic solution as individual generators' private data often contains confidential and sensitive information. Thus, the market design problem faces the ``privacy vs. effectiveness" dilemma.\footnote{In the economics literature (e.g., \cite{saari1978effective}), an \emph{effective} price mechanism refers to the one that converges to equilibrium prices for any choice of market participants' type. In this paper, we use the term ``effective" specifically to characterize a price mechanism that minimizes the social loss function defined by \eqref{eq:prob_0} below.}
\vspace{-2ex}
\subsection{Background}
\vspace{-2ex}
Market-based control is a common strategy to operate large-scale Cyber-Physical Systems (CPSs) such as power grids and transportation systems (\cite{ranter_dynamic_dual}). Unlike centralized control, market-based control relies on the spontaneous decisions of market participants (distributed controllers), whose behaviors are controlled indirectly by the market operator through incentives. Mechanism design, a subfield of economics, provides a theoretical basis for designing appropriate reward mechanisms. Market-based LFC has been studied from a mechanism design perspective previously by \cite{7798486} and \cite{ tanaka2012dynamic}. {\color{black}The former focuses on incentivizing generators to report the types truthfully, whereas the latter is concerned with incentivizing generators to implement distributed control law faithfully. The role of market-based LFC considered in this paper is closer to that of \cite{tanaka2012dynamic}.}

Since market-based control requires a significant amount of information exchange among participants, data confidentiality often becomes an issue. 
Indeed, when the participants consider the value of privacy to be greater than the incentive procured by traditional mechanism design, they may well behave against social benefit in an effort to protect their own privacy.  
Unfortunately, privacy has not been the main part of the conventional mechanism design framework, and this dilemma has been a well-recognized issue in the mechanism design research \cite[Chapter 10]{shoham2008multiagent}.


In order to address this privacy issue in the mechanism design literature, there have been attempts from various angles including 
differential privacy  (\cite{10.1145/2229012.2229073, kearns_pai_roth_ullman_2014}) and homomorphic encryption (HE) (\cite{gao2018privacy}).
In the meantime, 
HE has also been recognized as a solution to the privacy problem in networked control systems (\cite{7403296, FAROKHI2016163,KIM2016175}).
Straddling between these distinct research fields, this paper presents a solution to a privacy-preserving market design by perceiving it as a feedback control design problem and by borrowing HE tools from recent encrypted control systems literature.

\vspace{-2ex}
\subsection{Related work in encrypted control}
\vspace{-2ex}
\cite{7799042} and \cite{8262869} have demonstrated secure outsourcing of quadratic optimization in a multi-party computation framework. 
The latter especially showcased the use of HE combined with an additive blind. The work of \cite{8398387} and \cite{9030124} also considered the encryption of distributed controller problem but takes a different approach from our work. The major role of HE and other cryptographic primitives used therein is to ensure that during the evaluation of total control law, the neighboring agent's portion of the control gain (and the state) remains private to each other. In our work, the role of HE is more to assist the secure synthesis of a controller by protecting the confidentiality of each agent's cost functions. Our work does not need to use HE for hiding the controller evaluation because the local control portion already remains exclusive to the others. Recently, the pole-placement-based technique introduced by \cite{Kim_ARC} has enabled evaluating the encrypted dynamic controller in real-time without the bootstrapping operation \cite{homenc}, which is generally avoided for its computational overhead. 
\cite{9158986} utilized a controller reconstruction method to encrypt a nonlinear time-varying controller. \cite{Kim_NL} recently showed that systems admitting nonlinear observable canonical form can also be run for an infinite time horizon while encrypted. Despite recent advances, however, encryption of the dynamic market mechanism with HE is not developed yet.

\vspace{-1.5ex}
\subsection{Contribution}
\vspace{-1.5ex}
The main contribution of this paper is to eliminate the privacy constraint in the dynamic mechanism design problem using HE. We propose an encrypted market architecture that, by mechanism design, effectively redistributes some controller implementation effort to local clients via an incentive signal while also satisfying individual clients' preference for privacy.
The role of HE in our setting is somewhat different from that of the current paradigm of encrypted control, where the use of HE is to securely outsource the implementation of an already-synthesized controller. Instead, we use HE to implement distributed control laws while preserving privacy. We also point out that secure synthesis of control laws over HE is also an important component in the proposed setting.


\vspace{-1ex}
\section{Preliminaries}\label{prelim}
\vspace{-3ex}
\subsection{Notation}
\vspace{-2ex}
The set of real numbers and of integers are denoted with $\mathbb{R}$, and $\mathbb{Z}$, respectively. The set of integers modulo $q$ is denoted with $\mathbb{Z}_q$. The notation $||\cdot||_{W}$ means the weighted norm with a weight matrix $W$. We reserve notations $\hat{e}(\cdot)$ and $\hat{d}(\cdot)$ for the encryption and decryption operations of HE scheme and $\tilde{e}(\cdot)$ and $\tilde{d}(\cdot)$ for the other (non-homomorphic) encryption scheme. Likewise, $\hat{(\cdot)}$ or $\tilde{(\cdot)}$ over quantities mean that they are homomorphically or non-homomorphically encrypted variables, respectively.

\vspace{-2ex}
{\subsection{Price-based LFC}}
\vspace{-2ex}
Consider a linear time-invariant dynamics representing $N$ control areas for $i = 1, 2, \cdots, N$
\begin{subequations} \label{generator_ss}
\begin{alignat}{2}
        \label{stacked_system}
        x(t+1) &= Ax(t) + Bu(t) + B_{w}w(t) \\
         \label{indiv_system}
        x_i(t+1) &= \sum\nolimits_{j=1}^N A_{ij}x_j(t) + B_i u_i(t) + B_{w,i} w_i(t) \\
                 y_{i}(t) &= C_{i}x_{i}(t) + v_i(t)
\end{alignat}
\end{subequations}
where $x$, $u$, and $w$ are the stacked state, input, and disturbance (e.g., $x = [x_1^{\top}, x_2^{\top}, \cdots, x_{N}^{\top}]^{\top}$) and $y_{i}$ and $v_i$ are $i$-th generator's measurement output and the measurement noise, respectively. The system \eqref{generator_ss} is quite general but may be understood as a linearized swing equation in the context of LFC, and is further specified in Section \ref{simulation}.
\vspace{-1ex}
\subsubsection{Generator $i$'s optimal control problem:} Each generator $i$ is assumed to minimize its private instantaneous cost of operation which can be defined as:
\[
\ell_i(x_i(t), u_i(t))=\tfrac{1}{2}\|x_i(t)\|_{Q_i}^2+\tfrac{1}{2}\|u_i(t)\|_{R_i}^2.
\]
In addition, we consider a price mechanism that incentivizes the generator to produce adequate power. That is, at time step $t$, the $i$-th generator works to maximize its profit by selling the power at price $p(t)$, effectively solving
\begin{equation}
\label{eq:prob_i}
\min_{u_i}\limsup_{T\rightarrow \infty}
\tfrac{1}{T}\sum\nolimits_{k=t}^{t+T-1}\ell_i(x_i(k), u_i(k))-p(t)^\top y_i(k).
\end{equation}
{\color{black}Here, we assume that generators are not price-anticipating. That is, the price vector $p(t)$ announced by the ISO at time step $k=t$ will be assumed to be fixed for the generator's optimization until $k=t+T-1$.}
Since \eqref{eq:prob_i} is an LQ optimal control problem, the optimal control input $u_i(t)$ is written as a linear function of $x_i(t)$ and $p(t)$. That is,
\begin{equation}
\label{eq:ui_hat}
u_i(t)=\hat{u}_i(x_i(t), p(t))=F_i x_i(t)+M_i p(t)
\end{equation}
for some $F_i$ and $M_i$ (defined in Appendix \ref{appendix:a}).
In what follows, we define the collection of parameters $\theta_i := (A_{ii}, B_{i}, B_{w,i}, C_{i}, Q_i, R_i)$ as generator $i$'s type. We assume this information is not publicly known and contains sensitive information that generator $i$ wants to conceal.
\vspace{-1ex}
\subsubsection{Market operator's optimal control problem:} 
Anticipating generators' best responses \eqref{eq:ui_hat}, the objective of the market operator is to design a price mechanism that solves
\begin{align}
\min_p \; \limsup_{T\rightarrow \infty}\;  & \tfrac{1}{T}\sum\nolimits_{t=0}^{T-1}\Bigl(
\ell_0(x(t)) \nonumber \\
& +\sum\nolimits_{i=1}^N \ell_i(x_i(t), \hat{u}_i(x_i(t), p(t)))
\Bigr). \label{eq:prob_0}
\end{align}
A quadratic cost functional $
\ell_0(x(t))=\tfrac{1}{2}\|x(t)\|_{Q_0}^2
$ is assumed, and it ensures the quality of ancillary services. By solving \eqref{eq:prob_0}, the market operator minimizes the social loss (total generation cost). Since $\ell_j, j=0, 1, ..., N$ are quadratic, \eqref{eq:prob_0} is another LQ optimal control problem with the following market dynamics driven by the price
\begin{subequations}
\label{eq:closed_loop_market}
\begin{alignat}{2}
    x(t) &= A_p x(t) +B_p p(t) + B_{w}w(t) \\
    y(t) &= C_p x(t) + v_p(t) \\
    A_p  \! &:=\!\!\begin{bmatrix}
    (A_{11} \!+\! B_{1}F_{1})\!\!\!\! & \cdots \!\!\!\!\!\!\!\!\!\!\!\!\!\!\! & A_{1N} \\
    \vdots & \ddots & \vdots \\
    A_{N1} \!\!\! & \!\!\!\!\!\!\!\!\!\!\!\!\!\!\! \cdots & \!\!\!\!(A_{NN} \!+\! B_{N}F_{N})
    \end{bmatrix}\!\!, \; B_p\!:=\!\!\begin{bmatrix}
    B_{1} M_{1}\\
    \vdots\\
    B_{N} M_{N}
    \!\end{bmatrix} \nonumber  \\
    C_p \! &:=  \begin{bmatrix}C_{1} & \cdots & C_{N}\end{bmatrix}, v_p(t):=v_{1}(t) + \cdots + v_{N}(t) \nonumber
\end{alignat}
\end{subequations}

We use $Q$ to denote the total cost matrix, which will be a quadratic function of $Q_i$, $R_i$, $F_i, M_i$. This paper considers the output feedback setting, where the market operator decides the price input $p(t)$ of $\eqref{eq:closed_loop_market}$ based on a partial observation $y(t)$ of the grid state $x(t)$. 
Notice that solving the optimal control problem \eqref{eq:prob_0} becomes challenging for the market operator when the type vectors $\theta_i$ for $i = 1, \cdots, N$ are not fully known as generators do not wish to share them with the market operator.
{\color{black}
In the sequel, we assume generators report their \emph{encrypted} types truthfully; {\color{black}this is plausible due to the direct revelation principle (\cite{shoham2008multiagent}), where the participant's dominant strategy is truth-telling.}  The \emph{encrypted} optimal controller for \eqref{eq:prob_0} is then synthesized homomorphically based on the truthfully reported data. As is clear from the problem formulation above, the purpose of the optimal controller (price adjusting mechanism) is to incentivize generators to minimize the aggregated cost functional.}

\vspace{-1ex}
\subsection{Hybrid encryption method} \label{hybrid}
\vspace{-2ex}
We employ two different cryptosystems (one of them is homomorphic) to form a hybrid encryption scheme for our proposed architecture. For HE, we mainly adopt the FHE scheme called GSW-LWE cryptosystem (\cite{GSW}, \cite{GSW-LWE}) throughout this paper but the encrypted market architecture we propose is not limited to this particular scheme. Thus, we intend to keep it general and describe some core properties of the HE scheme that can serve the purpose of our work instead of prescribing the exact mechanics.

\vspace{-1ex}
\subsubsection{First cryptosystem (HE):}
\label{prelim:he}
Let the pair $(\text{sk1}, \text{pk1})$ denote secret and public keys for the first cryptosystem. The public keys are distributed to intended users prior to operations. Let $m \in \mathbb{Z}_q$ denote the plaintext, which is often the output of pre-processing some real-valued data by quantization. The modulo $q$ restricts the space of plaintext to the set of finite integers. The encryption operation uses the public key $\text{pk1}$ and it maps the plaintext $m$ to a ciphertext $c$ as $c := \hat{e}(m)$.
The decryption operation, given the secret key $\text{sk1}$, reverses (in some cases with a small ciphertext noise $\epsilon_1$) the encryption as
$m := \hat{d}(c) + \epsilon_1$.
For encrypted data, homomorphic addition and multiplication operations, in general, can be described as follows
$$
\textsf{EncAdd}(c_1, c_2) := \hat{e}(m_1) + \hat{e}(m_2) \mod q,
$$
$$
\hat{d}(\hat{e}(m_1) + \hat{e}(m_2) \mod q) = m_1 + m_2 + \epsilon_2,
$$
$$
\textsf{EncMult}(c_1, c_2) := \hat{e}(m_1) \cdot \hat{e}(m_2) \mod q,
$$
$$
\hat{d}(\hat{e}(m_1) \cdot \hat{e}(m_2) \mod q) = m_1 m_2 + \epsilon_3,
$$
with some ciphertext noises $\epsilon_2$ and $\epsilon_3$ that grow larger than $\epsilon_1$ as we apply more and more encrypted operations.
In addition, we may wish to have the bootstrapping operation. Bootstrapping (\cite{homenc}) is a ciphertext evaluation of the decryption function. This special operation refreshes the ciphertext noises generated from previous operations. This may assist the controller synthesis but we do not necessarily require this operation for our synthesis task.

\vspace{-1ex}
\subsubsection{Second cryptosystem:}
Let the pair $(\text{sk2}, \text{pk2})$ denote secret and public keys for the second cryptosystem. 
For this cryptosystem, which may not be a HE, we simply require encryption and decryption to satisfy the equation
$
\tilde{d}(\tilde{e}(m)) := m.
$
The role of the second cryptosystem is primarily to protect from the ISO eavesdropping the types being transmitted to the delegate server.

\begin{figure}[t] 
    \includegraphics[scale=0.55]{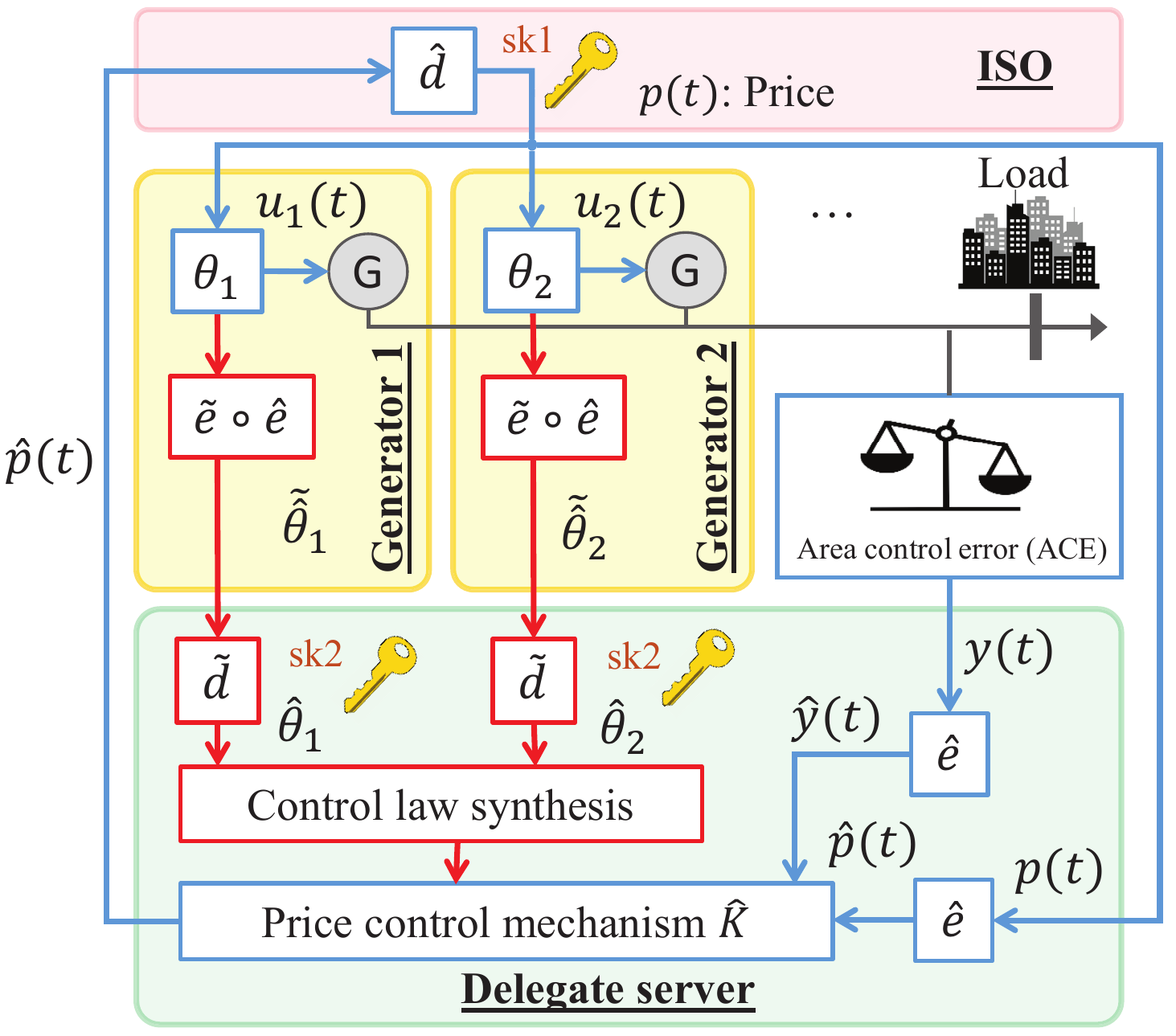}
    \caption{Frequency control in the encrypted market. Off-line (one-time) communication channels are colored red, and online (real-time) communication links in blue.}
    \label{fig:market}
\vspace{-1ex}
\end{figure}

\vspace{-1ex}
\section{Encrypted Market Mechanism}
\vspace{-2ex}
\subsection{Encrypted market architecture} \label{architecture}
\vspace{-2ex}
The proposed market architecture is shown in Fig.~\ref{fig:market}. Throughout this paper, we assume that collusion between the market operator (ISO) and the delegate server is illegal. In the adversarial setting, both the ISO and the delegate server can be considered benign passive attackers; they are curious about data but follow the protocol faithfully (semi-honest). Privacy is protected by a hybrid encryption method, where two separate cryptosystems described above are combined. To achieve secrecy in control synthesis, we proposed introducing a delegate server, while the second cryptosystem, on top of the first, provides secrecy in the transmission of private types. The server also continuously operates the encrypted price mechanism (encrypted controller evaluation) on behalf of the ISO, which is made possible by using the first cryptosystem (HE).
Prior to the market operation, the ISO generates a pair $(\text{sk1}, \text{pk1})$ of secret and public keys, and the latter is published. The second cryptosystem is a public-key encryption system (which does not need to support homomorphic operations) and is managed by the delegate server. The delegate server generates a pair $(\text{sk2}, \text{pk2})$
and the latter is published. The operation of the proposed market mechanism is divided into (A) the off-line phase (one-time operations, no real-timeness required) and (B) the online phase (continuous, real-time operation):
\begin{enumerate}[label=\quad(\Alph*), wide=1pt]
\vspace{-1ex}
    \item Off-line phase: In this phase, the delegate server collects generators' (encrypted) type vectors $\theta_i$ and designs an optimal price adjusting mechanism (control law) that minimizes the social loss. Each generator first encrypts $\theta_i$ using pk1 (the corresponding encryption operation is denoted by $\hat{e}$) to generate $\hat{\theta}_i=\hat{e}(\theta_i)$, and then further encrypts the output using pk2 (the corresponding encryption function is $\tilde{e}$) to generate $\tilde{\hat{\theta}}_i=\tilde{e}(\hat{\theta}_i)$. Notice that the function $\tilde{e}$ can be reversed only by the delegate server using sk2 to perform the decryption $\tilde{d}$. Based on $\hat{\theta}_i, i=1,2, \cdots, N$, the server synthesizes a control law $\hat{K}$ homomorphically. The synthesis process, if necessary, can be aided by the ISO while still preserving privacy.
    
    \item Online phase: In this phase, the delegate server continuously monitors the output signal $y(t)$ and keeps computing an (encrypted) price vector $\hat{p}(t)$ to suppress the frequency deviation. Since the server only knows the encrypted controller $\hat{K}$, $y(t)$ is first encrypted using pk1 so that $\hat{y}(t)=\hat{e}(y(t))$ can be processed by $\hat{K}$. The controller output $\hat{p}(t)$ is transmitted to the ISO where the price vector is decrypted as $p(t)=\hat{d}(\hat{p}(t))$. The price $p(t)$ is announced to the generators, who adjust command inputs $u_i(t)$ to the local turbines in response to the real-time price. In order for the price adjusting mechanism to be sufficiently faster than the time scale of the underlying swing equation, the feedback control loop must be ``closed" in the order of seconds. Therefore, the controller $\hat{K}$ must be operated in real-time, and this may render the use of computationally expensive operations such as bootstrapping prohibitive.
\end{enumerate}
\vspace{-1ex}
Note that the effectiveness of the proposed encrypted price mechanism can be made equal to that of the unencrypted counterpart by the proper choice of encryption parameters. However, because of the hybrid encryption, both the ISO and the delegate server (assuming they do not share their secret keys) do not learn anything more than the price $p(t)$ and the output $y(t)$. Furthermore, generator $i$'s type $\theta_i$ is never learned by any party in the system, even if the communication channels between different parties in Fig.~\ref{fig:market} are eavesdropped by any unauthenticated agents.

\vspace{-1.5ex}
\subsection{Encrypted controller synthesis and evaluation}
\vspace{-2ex}
Controller synthesis generally involves highly nonlinear and recursive operations. Recall (Section \ref{hybrid}) that the ciphertext noise can grow with encrypted operations, which can lead to incorrect decryption. This makes bootstrapping the ciphertext useful and often necessary as it refreshes the ciphertext noise level. However, bootstrapping in general can be expensive and impractical in some cases with high circuit complexities. For example, matrix inversion appears frequently in controller synthesis problems but it is not so straightforward to compute over HE.

Alternatively, the delegate server may avoid bootstrapping by collaborating with the ISO (who has sk1) to form a two-party protocol for a secure matrix inversion (Algorithm 1) and a masked re-encryption (Algorithm 2). The idea of Algorithm 1 is borrowed from \cite{MIC}; it allows the ISO to invert a matrix on behalf of the delegate server while the data content is kept secret by a secure transformation-based masking technique (with no repetition of the same transformation). In a similar vein, Algorithm 2 allows the ISO to refresh the ciphertext noise level for the delegate server.
These protocols may be used as described below to solve algebraic Riccati equations and to perform pole placement in the ciphertext domain.

\vspace{-1.5ex}
\begin{algorithm}[H]
 \caption{Encrypted matrix inversion}
\vspace{-0.2ex}
The encrypted $\hat{M}$ is inverted without revealing $M$.\\
$\textsf{Delegate server}:$ Construct invertible transformation matrices $\Phi_1$ and $\Phi_2$. Obtain ciphertexts $\hat{\Phi}_i = \hat{e}(\Phi_i)$ and $\hat{\Phi}^{-1}_i = \hat{e}(\Phi^{-1}_i)$ for $i=1,2$. Compute the masked encrypted matrix $\hat{\tilde{M}} = \hat{\Phi}_1 \hat{M}\hat{\Phi}_2$ by transformation over HE. \\
\begin{tabular}{@{} l @{}} \end{tabular}
\vspace{-0.2ex}
\quad \textsf{Return to ISO:} $\hat{\tilde{M}}$\vspace{-0.2ex}\\
$\textsf{ISO}:$ Decrypt the ciphertext $\hat{\tilde{M}}$ with \text{sk1} to obtain $\tilde{M} := \Phi_1 M \Phi_2$. Invert $\tilde{M}$ to obtain $\tilde{M}^{-1} = \Phi^{-1}_2M^{-1}\Phi^{-1}_1$. Encrypt the result and get $\hat{\tilde{M}}^{-1} = \hat{e}(\tilde{M}^{-1})$.\\
\begin{tabular}{@{} l @{}} \end{tabular}
\vspace{-0.5ex}
\quad \textsf{Return to Delegate server :} $\hat{\tilde{M}}^{-1}$\vspace{-0.2ex}\\
$\textsf{Delegate server}:$ Unmask the ciphertext $\hat{\tilde{M}}^{-1}$ by multiplying over HE to acquire $\hat{M}^{-1} = \hat{\Phi}_2\hat{\tilde{M}}^{-1}\hat{\Phi}_1$.\\
\begin{tabular}{@{} l @{}} \end{tabular}
\quad \textsf{Return:} $\hat{M}^{-1}$\\
\vspace{-2.5ex}
\label{alg:enc_matrix_inversion}
\end{algorithm}
\vspace{-2.9ex}
\begin{algorithm}[H]
 \caption{Masked re-encryption}
The encrypted $\hat{M}$ is re-encrypted without revealing $M$.\\
$\textsf{Delegate server}:$ Construct invertible transformation matrices $\Phi_1$ and $\Phi_2$. Obtain ciphertexts $\hat{\Phi}_i = \hat{e}(\Phi_i)$ and $\hat{\Phi}^{-1}_i = \hat{e}(\Phi^{-1}_i)$ for $i=1,2$. Compute the masked encrypted matrix $\hat{\tilde{M}} = \hat{\Phi}_1 \hat{M}\hat{\Phi}_2$ by transformation over HE. \\
\begin{tabular}{@{} l @{}} \end{tabular}
\quad \textsf{Return to ISO:} $\hat{\tilde{M}}$\\
$\textsf{ISO}:$ Decrypt the ciphertext $\hat{\tilde{M}}$ with \text{sk1} to obtain $\tilde{M}$. Encrypt again to produce $\hat{\tilde{M}} = \hat{e}(\tilde{M})$ where $\tilde{M} := \Phi_1 M \Phi_2$.\\
\begin{tabular}{@{} l @{}} \end{tabular}
\quad \textsf{Return to Delegate server :} $\hat{\tilde{M}}$\vspace{-0.5ex}\\
$\textsf{Delegate server}:$ Unmask the ciphertext $\hat{\tilde{M}}$ by multiplying over HE to acquire $\hat{M} = \hat{\Phi}^{-1}_1\hat{\tilde{M}}\hat{\Phi}^{-1}_2$.\\
\begin{tabular}{@{} l @{}} \end{tabular}
\quad \textsf{Return:} $\hat{M}$\\
\label{alg:pseudo_bootstrapping}
\vspace{-2ex}
\end{algorithm}
\vspace{-4ex}
\subsubsection{Encrypted algebraic Riccati equation:}
We consider the encrypted synthesis of $\mathcal{H}_2$ optimal controller $\hat{K}= (\hat{A}_K, \hat{B}_K, \hat{C}_K, \hat{D}_K)$. The synthesis requires the encrypted aggregation of type vectors $\theta = (\theta_1, \cdots, \theta_N)$ at the server. This contains the encrypted system information, e.g., $\hat{A}_p$, $\hat{B}_p$, $\hat{B}_w$, and $\hat{C}_p$ of \eqref{eq:closed_loop_market} along with the encrypted total cost weights $\hat{Q}$ and $\hat{R}$. Solving discrete-time algebraic Riccati equations (DARE) is to find the unique positive semi-definite solutions $\hat{P_1}$ and $\hat{P_2}$ to the following quadratic matrix equations over ciphertexts
\[
\hat{P}_1=\hat{A}_{p}^{\top}\hat{P}_1\hat{A}_{p}-(\hat{A}_{p}^{\top}\hat{P}_1\hat{B}_p)(\hat{R}+\hat{B}_{p}^{\top}\hat{P}_1\hat{B}_{p})^{-1}(\hat{B}^{\top}_{p}\hat{P}_1\hat{A}_{p})+\hat{Q},
\]
\[
\hat{P_2}=\hat{A}_{p}\hat{P}_2\hat{A}_{p}^{\top}-(\hat{A}_{p}\hat{P}_2\hat{C}_{p}^{\top})(\hat{V}+\hat{C}_{p}\hat{P}_2\hat{C}_{p}^{\top})^{-1}(\hat{C}_{p}\hat{P}_2\hat{A}_{p}^{\top})+\hat{W}.
\]
where $\hat{V}$ and $\hat{W}$ are the appropriate measurement and disturbance covariance matrices.
The controller and observer gains can be recovered from
\[\hat{K}_p := (\hat{R}+\hat{B}_{p}^{\top}\hat{P}_1\hat{B}_{p})^{-1}(\hat{B}^{\top}_{p}\hat{P}_1\hat{A}_{p}),\]
\[\hat{L}_p :=(\hat{A}_{p}\hat{P}_2\hat{C}_{p}^{\top})(\hat{V}+\hat{C}_{p}\hat{P}_2\hat{C}_{p}^{\top})^{-1}.\]
A standard recursive method can be performed off-line using tools such as a standard bootstrapping technique or Algorithms \ref{alg:enc_matrix_inversion} and \ref{alg:pseudo_bootstrapping} presented below. Note that other algorithms such as Newton's method and doubling algorithm (\cite{guo_lin_xu_2006}) or a direct generalized eigenvalue-based algorithm (\cite{pappas1980numerical}) exist and their performance should be compared in terms of accuracy and efficiency for ciphertext implementations. 
The resulting Riccati-based controller is written as follows
\begin{subequations}
\begin{alignat}{2} \label{price_controller}
\hat{x}_K(t+1) & = \hat{A}_K \hat{x}_K + \hat{B}_K \hat{y}(t) + \hat{B}_w \hat{w}(t), \\
    \hat{p}(t) & = \hat{C}_K \hat{x}_K(t),
\end{alignat}
\end{subequations}
where $\hat{A}_K := \hat{A}_p - \hat{L}_p \hat{C}_{p} - \hat{B}_p \hat{K}_p$, $\hat{B}_K := \hat{L}_p$, $\hat{C}_K := -\hat{K}_p$.
\vspace{-4ex}
\subsubsection{Encrypted pole-placement:}
Once $\hat{K}$ is synthesized, the ISO and the server must ensure that the synthesized controller can operate in real-time. This amounts to running the linear dynamic controller \eqref{price_controller} over the encrypted state $\hat{x}_K$, the encrypted plant output $\hat{y}$, and the encrypted disturbance $\hat{w}(t)$. In order to avoid costly operations (e.g., decrypting the entire internal state every time, or frequent bootstrapping), it is known that the state matrix $\hat{A}_K$ must already consist of integer values without having been quantized (\cite{integer_matrix}). Since the controller we obtain likely consists of non-integer values (when decrypted), we need to find an integer approximation. This can be achieved by a pole-placement-based transformation technique introduced in \cite{Kim_TAC}. But prior to that, it is important to note that the unencrypted pair $(A_K, C_K)$ must be an observable pair for pole placement to be possible.\footnote{A minimal realization of the controller satisfies the observability.} Thus, the synthesis and processing of the controller for real-time operation consist of first ensuring the observability by constructing and inverting the observability matrix, followed by performing the pole-placement-based transformation over HE. Ackermann's formula is a standard pole-placement technique, which requires evaluating the inverse of the observability matrix and the matrix characteristic polynomial and can be efficiently implemented over HE using Algorithms 1 and 2.
\vspace{-2ex}
\subsubsection{Quantization and encryption parameters:} Quantization has been known to be important in guaranteeing the performance of an encrypted controller. Given the security requirement specified through $\lambda$-bit security parameter, a detailed guideline for choosing the encryption (and quantization) parameters ensuring the equal performance of the encrypted controller can be found in the literature (\cite{9304509}). Adopting the same notations, we should select the parameters $\textsf{scale}:= (L,s_1,s_2,r)$ with $s_2 = 1$ if the realized system is single-input, single-output (SISO). This then leads to selecting the appropriate encryption parameters $\textsf{param} = (q, \sigma, n_{L}, \nu, d)$. The resulting encrypted controller $\hat{K}$ will then be ready to be used for the online phase specified in Section \ref{architecture}-(B). \label{parameters}
\vspace{-2ex}
\section{Numerical simulation}
\vspace{-1ex}
\label{simulation}
\begin{table}[htbp]
\vspace{-2ex}
\caption{System parameters (\cite{powersystembook})}
\vspace{-1ex}
\begin{tabular}{|p{2.0cm}||p{3.0cm}|p{0.90cm}|p{0.90cm}| }
\hline
     Parameters & Definition & Area 1 & Area 2\\
     \hline
     $H \; (pu \cdot s)$ & inertia constant  & 0.081 & 0.091\\
     $D \; (pu/Hz)$ & damping coefficient & 0.015 & 0.015\\
     $T_{ij} (pu/Hz) $ & tie-line coefficient & 0.20 & 0.20 \\
     $R \; (Hz/pu)$ & droop characteristic  & 3.100 & 2.631 \\
     $T_{g} \; (s)$ & governor time constant & 0.070 & 0.067 \\
     $T_{t} \; (s)$ & turbine time constant & 0.393 & 0.387 \\
     \hline
\end{tabular}
\vspace{-1ex}
\label{power_parameters}
\end{table}
\vspace{-1ex}
The simplified swing equations given below are adopted from the $n$-generators, $N$-control area LFC model from \cite{powersystembook} as a case study. We consider a case with two areas ($N=2$) with each area containing one generator ($n=1$) with system parameters listed in Table \ref{power_parameters}.
\begin{subequations} \label{swing}
\begin{alignat}{2}
        \tfrac{d\Delta f_{i}}{dt} & = -\tfrac{D_i}{2H_i}\Delta f_{i} - \tfrac{\Delta P_{{tie,i}}}{2H_i} + \tfrac{\Delta P_{{m}_{i}}}{2H_i} -\tfrac{\Delta P_{L_i}}{2H_i}\\ 
        \tfrac{d\Delta P_{{tie,i}}}{dt} & = (2\pi T_{ij})\Delta f_{i} - (2\pi T_{ij}) \Delta f_{j} \label{coupling}\\ 
        \tfrac{d\Delta P_{m,i}}{dt} & = -\tfrac{\Delta P_{m,i}}{T_{t,i}} +\tfrac{\Delta P_{g,i}}{T_{t,i}} \\ 
        \tfrac{d\Delta P_{g,i}}{dt} & = -\tfrac{\Delta f_{i}}{(T_{g,i}R_{i})} - \tfrac{\Delta P_{g, i}}{T_{t,i}} + \tfrac{\Delta P_{c,i}}{T_{g,i}} \\ 
        \text{ACE}_{i} & = (\tfrac{1}{R} + D)_{i} \Delta f_{i} + \Delta P_{{tie,i}},
\end{alignat}
\end{subequations}
where $\Delta f$ is the frequency deviation, $\Delta P_{tie}$ is the net tie-line power flow, $\Delta P_m$ is the change in mechanical power, $\Delta P_g$ is the change in governor valve position, $\Delta P_c$ is the control input to the plant, $\Delta P_L$ is the load change disturbance (electricity on-demand), and the area control error ($\text{ACE}$) is a standard output in the power system.

\begin{figure}[t]
    \centering
    \includegraphics[scale=0.36, clip]{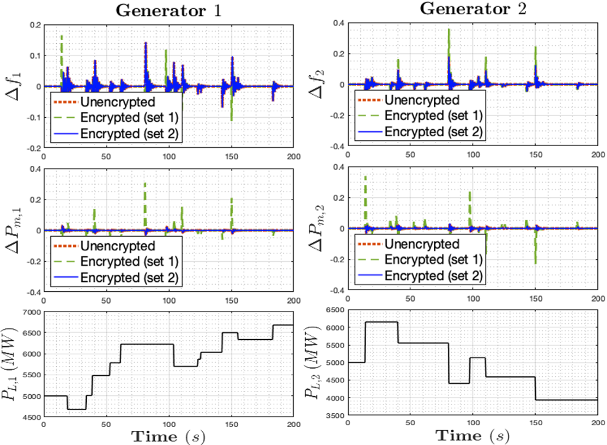}
    \vspace{-2ex}
    \caption{Frequency control under the encrypted price.}
    \vspace{-1ex}
    \label{fig:lfc}
\end{figure}

\begin{figure}[t]
    \centering
    \vspace{-1ex}
    \includegraphics[scale=0.37, trim={0.1cm 0.1cm 0.01cm 0.1cm},clip]{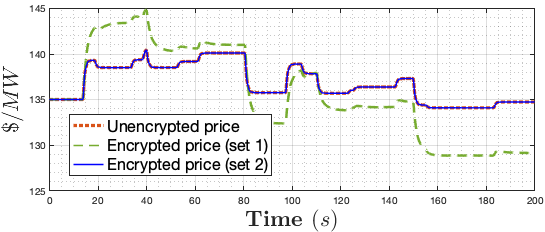}
    \vspace{-2ex}
    \caption{The price $p(t)$ of the encrypted controller on $\Delta P_m$.}
    \label{fig:price}
\end{figure}

With $x_{i}^{\top}(t) = \begin{bmatrix}
\Delta f_{i} & \Delta P_{tie,i} & \Delta P_{m,i} & \Delta P_{g,i}
\end{bmatrix}$ as the state of $i$-th control area generator, the state-space model conforming to \eqref{generator_ss} is discretized with sampling time $0.2\:s$.
The private cost matrices used for each generator were $Q_1=\text{diag}(100, 50, 20, 500)$, $R_1=100$ and $Q_2=\text{diag}(200, 45, 30, 400)$, $R_2=50$, respectively. The market operator also had a social cost matrix $Q_0 = \text{diag}(600, 50, 50, 20, 400, 120, 50, 30)$ and $R_0 = 25$.
Since the generated electrical power from the $i$-th control area is closely related to the mechanical power, $\Delta P_{m,i}$, the price $p(t)$ is designed to incentivize the generation of power when the load demand changes. For each generator area, random load changes were sampled (with the baseline load at $5000 \text{MW}$). 
Frequency regulation results and the corresponding price response $p(t)$ (the baseline price at $\$135/\text{MW}$) are plotted in Fig. \ref{fig:lfc}, and Fig. \ref{fig:price}, respectively.
Simulated results show two encrypted price graphs based on two different pairs of parameters set $\textsf{scale}$ and $\textsf{params}$ discussed in \ref{parameters}) and their corresponding trajectories. Each pair was obtained by first setting the security parameter to $\lambda = 32$ bits and choosing the scale parameter $L$ (per guideline in \cite{9304509}). The first set (dashed green) was obtained with $L = 2^{-6}$, resulting in $\textsf{scale}_1 = (2^{-6}, 2^{-12}, 1, 2^{-12})$ and $\textsf{param}_1 = (2^{30}, 1, 329, 2, 19)$ but the result shows that this pair is not sufficient in suppressing the encryption noise. The second set (blue) was obtained with a smaller value of $L = 2^{-10}$, resulting in $\textsf{scale}_2 = (2^{-10}, 2^{-20}, 1, 2^{-20})$ and $\textsf{param}_2 = (2^{60}, 1, 648, 2, 35)$, and it achieved the performance equal to the unencrypted. The result first validates the feasibility of the proposed encrypted market mechanism, and also reinforces the importance of choosing encryption and quantization parameters in implementation.
\vspace{-1ex}
\section{Conclusion and future work}
\vspace{-1ex}
We proposed an encrypted market architecture that fulfills a privacy-preserving market mechanism. The privacy issue is ubiquitous on many fronts of technological advancement. For this reason, realizing a practical privacy-aware market mechanism may find many practical uses beyond the power system applications shown in this work. Future work may complement the present paper by further examining the security and practicality of techniques discussed in this paper for the off-line processes. This includes, for example, detailed implementations of encrypted controller synthesis, e.g., the encrypted Riccati equation, or pole placement. Conceptually, this paper integrated the theory of mechanism design with encrypted control to tackle the privacy problem. However, it would also be interesting to examine other aspects of CPS security besides privacy.
\vspace{-1ex}
\bibliography{ifacconf}

\begin{thebibliography}{29}
\providecommand{\natexlab}[1]{#1}
\providecommand{\url}[1]{\texttt{#1}}
\providecommand{\urlprefix}{URL }
\expandafter\ifx\csname urlstyle\endcsname\relax
  \providecommand{\doi}[1]{doi:\discretionary{}{}{}#1}\else
  \providecommand{\doi}{doi:\discretionary{}{}{}\begingroup
  \urlstyle{rm}\Url}\fi

\bibitem[{Alexandru et~al.(2017)Alexandru, Gatsis, and Pappas}]{8262869}
Alexandru, A.B., Gatsis, K., and Pappas, G.J. (2017).
\newblock Privacy preserving cloud-based quadratic optimization.
\newblock In \emph{55th Allerton Conference}.

\bibitem[{Alexandru et~al.(2019)Alexandru, Schulze~Darup, and Pappas}]{9030124}
Alexandru, A.B., Schulze~Darup, M., and Pappas, G.J. (2019).
\newblock Encrypted cooperative control revisited.
\newblock In \emph{2019 IEEE 58th CDC}, 7196--7202.

\bibitem[{Berger and Schweppe(1989)}]{berger1989real}
Berger, A.W. and Schweppe, F.C. (1989).
\newblock Real time pricing to assist in load frequency control.
\newblock \emph{IEEE Transactions on Power Systems}, 4(3), 920--926.

\bibitem[{Bevrani(2009)}]{powersystembook}
Bevrani, H. (2009).
\newblock \emph{Robust Power System Frequency Control}.
\newblock Springer.

\bibitem[{Cheon et~al.(2018)Cheon, Han, Kim, Kim, and Shim}]{integer_matrix}
Cheon, J.H., Han, K., Kim, H., Kim, J., and Shim, H. (2018).
\newblock Need for controllers having integer coefficients in homomorphically
  encrypted dynamic system.
\newblock In \emph{2018 IEEE CDC}.

\bibitem[{Chillotti et~al.(2018)Chillotti, Gama, Georgieva, and
  Izabachène}]{GSW-LWE}
Chillotti, I., Gama, N., Georgieva, M., and Izabachène, M. (2018).
\newblock {TFHE}: Fast fully homomorphic encryption over the torus.
\newblock Cryptology ePrint Archive.

\bibitem[{Farokhi et~al.(2016)Farokhi, Shames, and Batterham}]{FAROKHI2016163}
Farokhi, F., Shames, I., and Batterham, N. (2016).
\newblock Secure and private cloud-based control using semi-homomorphic
  encryption.
\newblock \emph{6th IFAC NECSYS}.

\bibitem[{Gao et~al.(2018)Gao, Yu, Liang, Hatcher, and Lu}]{gao2018privacy}
Gao, W., Yu, W., Liang, F., Hatcher, W.G., and Lu, C. (2018).
\newblock Privacy-preserving auction for big data trading using homomorphic
  encryption.
\newblock \emph{IEEE Transactions on Network Science and Engineering}, 7(2),
  776--791.

\bibitem[{Gentry(2009)}]{homenc}
Gentry, C. (2009).
\newblock \emph{A fully homomorphic encryption scheme}.
\newblock Ph.D. thesis, Stanford University.

\bibitem[{Gentry et~al.(2013)Gentry, Sahai, and Waters}]{GSW}
Gentry, C., Sahai, A., and Waters, B. (2013).
\newblock Homomorphic encryption from learning with errors:
  Conceptually-simpler, asymptotically-faster, attribute-based.

\bibitem[{Guo et~al.(2006)Guo, Lin, and Xu}]{guo_lin_xu_2006}
Guo, X.X., Lin, W.W., and Xu, S.F. (2006).
\newblock A structure-preserving doubling algorithm for nonsymmetric algebraic
  riccati equation.
\newblock \emph{Numerische Mathematik}, 103(3).

\bibitem[{Kearns et~al.(2014)Kearns, Pai, Roth, and
  Ullman}]{kearns_pai_roth_ullman_2014}
Kearns, M., Pai, M.M., Roth, A., and Ullman, J. (2014).
\newblock Mechanism design in large games: Incentives and privacy.
\newblock \emph{American Economic Review}, 104(5), 431–435.

\bibitem[{Kim et~al.(2021)Kim, Farokhi, Shames, and Shim}]{Kim_NL}
Kim, J., Farokhi, F., Shames, I., and Shim, H. (2021).
\newblock Toward nonlinear dynamic control over encrypted data for infinite
  time horizon.

\bibitem[{Kim et~al.(2022{\natexlab{a}})Kim, Kim, Song, Shim, Sandberg, and
  Johansson}]{Kim_ARC}
Kim, J., Kim, D., Song, Y., Shim, H., Sandberg, H., and Johansson, K.H.
  (2022{\natexlab{a}}).
\newblock Comparison of encrypted control approaches and tutorial on dynamic
  systems using learning with errors-based homomorphic encryption.
\newblock In \emph{Annual Reviews in Control}, volume~54, 200--218.

\bibitem[{Kim et~al.(2016)Kim, Lee, Shim, Cheon, Kim, Kim, and
  Song}]{KIM2016175}
Kim, J., Lee, C., Shim, H., Cheon, J.H., Kim, A., Kim, M., and Song, Y. (2016).
\newblock Encrypting controller using fully homomorphic encryption for security
  of cyber-physical systems.
\newblock \emph{IFAC-PapersOnLine}.
\newblock 6th IFAC NECSYS.

\bibitem[{Kim et~al.(2020)Kim, Shim, and Han}]{9304509}
Kim, J., Shim, H., and Han, K. (2020).
\newblock Design procedure for dynamic controllers based on {LWE}-based
  homomorphic encryption to operate for infinite time horizon.
\newblock In \emph{59th IEEE CDC}, 5463--5468.

\bibitem[{Kim et~al.(2022{\natexlab{b}})Kim, Shim, and Han}]{Kim_TAC}
Kim, J., Shim, H., and Han, K. (2022{\natexlab{b}}).
\newblock Dynamic controller that operates over homomorphically encrypted data
  for infinite time horizon.
\newblock \emph{IEEE TAC}.

\bibitem[{Kogiso and Fujita(2015)}]{7403296}
Kogiso, K. and Fujita, T. (2015).
\newblock Cyber-security enhancement of networked control systems using
  homomorphic encryption.
\newblock In \emph{2015 54th IEEE CDC}, 6836--6843.

\bibitem[{Lei et~al.(2013)Lei, Liao, Huang, Li, and Hu}]{MIC}
Lei, X., Liao, X., Huang, T., Li, H., and Hu, C. (2013).
\newblock Outsourcing large matrix inversion computation to a public cloud.
\newblock \emph{IEEE Transactions on Cloud Computing}.

\bibitem[{Nissim et~al.(2012)Nissim, Orlandi, and
  Smorodinsky}]{10.1145/2229012.2229073}
Nissim, K., Orlandi, C., and Smorodinsky, R. (2012).
\newblock Privacy-aware mechanism design.
\newblock In \emph{Proceedings of the 13th ACM Conference on Electronic
  Commerce}. Association for Computing Machinery, New York, USA.

\bibitem[{Pappas et~al.(1980)Pappas, Laub, and Sandell}]{pappas1980numerical}
Pappas, T., Laub, A., and Sandell, N. (1980).
\newblock On the numerical solution of the discrete-time algebraic riccati
  equation.
\newblock \emph{IEEE TAC}, 25(4), 631--641.

\bibitem[{Rantzer(2009)}]{ranter_dynamic_dual}
Rantzer, A. (2009).
\newblock Dynamic dual decomposition for distributed control.
\newblock In \emph{2009 American Control Conference}.

\bibitem[{Saari and Simon(1978)}]{saari1978effective}
Saari, D.G. and Simon, C.P. (1978).
\newblock Effective price mechanisms.
\newblock \emph{Econometrica}, 1097--1125.

\bibitem[{Schulze~Darup et~al.(2019)Schulze~Darup, Redder, and
  Quevedo}]{8398387}
Schulze~Darup, M., Redder, A., and Quevedo, D.E. (2019).
\newblock Encrypted cooperative control based on structured feedback.
\newblock \emph{IEEE Control Systems Letters}, 3(1).

\bibitem[{Shoham and Leyton-Brown(2008)}]{shoham2008multiagent}
Shoham, Y. and Leyton-Brown, K. (2008).
\newblock \emph{Multiagent systems: Algorithmic, game-theoretic, and logical
  foundations}.
\newblock Cambridge University Press.

\bibitem[{Shoukry et~al.(2016)Shoukry, Gatsis, Alanwar, Pappas, Seshia,
  Srivastava, and Tabuada}]{7799042}
Shoukry, Y., Gatsis, K., Alanwar, A., Pappas, G.J., Seshia, S.A., Srivastava,
  M., and Tabuada, P. (2016).
\newblock Privacy-aware quadratic optimization using partially homomorphic
  encryption.
\newblock In \emph{IEEE 55th CDC}.

\bibitem[{Tanaka et~al.(2012)Tanaka, Cheng, and Langbort}]{tanaka2012dynamic}
Tanaka, T., Cheng, A.Z.W., and Langbort, C. (2012).
\newblock A dynamic pivot mechanism with application to real time pricing in
  power systems.
\newblock In \emph{2012 American Control Conference}.

\bibitem[{Tanaka and Gupta(2016)}]{7798486}
Tanaka, T. and Gupta, V. (2016).
\newblock Incentivizing truth-telling in mpc-based load frequency control.
\newblock In \emph{2016 IEEE 55th CDC}.

\bibitem[{Teranishi et~al.(2020)Teranishi, Kogiso, and Ueda}]{9158986}
Teranishi, K., Kogiso, K., and Ueda, J. (2020).
\newblock Encrypted feedback linearization and motion control for manipulator
  with somewhat homomorphic encryption.
\newblock In \emph{IEEE/ASME International Conference on AIM}.

\end{thebibliography}
\vspace{-1ex}

\appendix
\section{Generator's control law}
\label{appendix:a}
The optimal control input $u_{i}(k)$ for \eqref{eq:ui_hat} can be written as $u_{i}(k) = F_{i}(k)x_{i}(k) + G_{i}(k)r_{i}(k+1)$, where
\begin{subequations} \label{eq:gains}
\begin{align}
    S_{i}(k) & = Q_{i} + A_{ii}^{\top}S_{i}(k+1)A_{ii} - A_{ii}^{\top}S_{i}(k+1)B_{i} \nonumber \\
    &(R_{i} + B_{i}^{\top}S_{i}(k+1)B_{i})^{-1}B_{i}^{\top}S_{i}(k+1)A_{ii}, \\
    S_i(T) & = Q_i, \\
    G_i(k) & = -(R_i + B_{i}^{\top}S_{i}(k+1)B_{i})^{-1}B_{i}^{\top}, \\
    F_i(k) & = G_{i}(k)S_{i}(k+1)A_{ii}, \\
    r_{i}(k) & = -C^{\top}p(t) + A_{ii}^{\top}(I - S_{i}(k+1)B_{i}(R_{i} \nonumber \\
    & + B_{i}^{\top}S_{i}(k+1)B_{i})^{-1}B_{i}^{\top}r_{i}(k+1), \\
    r_{i}(T) & = I.
\end{align}
for $k = T-1, \cdots, 1$, and $i = 1, \cdots, N$. Moreover, the steady-state solutions $\bar{S}_{i}$, and $\bar{r}_{i}$ exist, so that the optimal control law is
$u_{i}(t) = F_{i}x_{i}(t) + M_{i}p(t)$, where 
\begin{align}
    \bar{G}_{i} & = -(R_i + B_{i}^{\top}\bar{S}_{i}B_{i})^{-1}B_{i}^{\top}\\
    F_{i} & = \bar{G}_{i}\bar{S}_{i}A_{ii}, \\
    \Phi_{i} & = -(I - A_{ii}^{\top} + A_{ii}^{\top}\bar{S}_{i}B_{i}(R_{i} \nonumber \\
    & + B_{i}^{\top}\bar{S}_{i}B_{i})^{-1}B_{i}^{\top})^{-1}C^{\top}, \\
    \bar{r}_{i} & = \Phi_{i}p(t), \\
    M_{i} & = \bar{G}_{i}\Phi_{i}.
\end{align}

\end{subequations}



\end{document}